\newcommand{\be}{\begin{equation}}
\newcommand{\ee}{\end{equation}}
\newcommand{\ba}{\begin{array}{c}}
\newcommand{\ea}{\end{array}}
\newcommand{\bqa}{\begin{eqnarray}}
\newcommand{\eqa}{\end{eqnarray}}
\documentclass[
   ,final
  ]
  {aipproc}

\layoutstyle{6x9}
\SetInternalRegister\hbadness{8000} % pseudo latin isn't breaking very well :-)

\newcommand\doingARLO[2][]{%
  \ifx\mmref\undefined #1\else #2\fi
}

\begin{document}

\title
      {On the nature of the lightest scalar resonances\footnote{Talk presented by Zheng at ``Quark Confinement
      and Hadron Spectroscopy VI'', 21--25 Sept. 2004,  Cagliari, Italy} }


\author{Z.~X.~Sun, L.~Y.~Xiao, Z.~G.~Xiao, H.~Q.~Zheng and Z.~Y. Zhou}{
  address={Department of Physics, Peking University, Beijing 100871, P.~R.~China},
}

% \copyrightholder{Acoustical Scociety of America}
\copyrightyear  {2001}

\begin{abstract}
We briefly review the recent progresses in the new unitarization
approach being developed by us. Especially we discuss the large
$N_c$ $\pi\pi$ scatterings  by making
 use of the partial wave $S$ matrix parametrization form. We find that
 the $\sigma$ pole may move to
the negative real axis on the second sheet of the complex $s$
plane, therefore it raises the interesting question that this
`$\sigma$' pole may be related to the $\sigma$ in the linear
$\sigma$ model.
\end{abstract}

\date{\today}
\maketitle

The problem of how to restore unitarity and meanwhile respecting
chiral perturbation amplitudes at low energies is very interesting
and also difficult. A simple solution one has when dealing with
such a difficult problem is the Pad\'e approximation and its
variations, which achieved some phenomenological success.
Nevertheless, the Pad\'e approximation encounters serious
problems~\cite{Qin} which can hardly be resolved within the method
itself. For this reason, it is worthwhile to make further efforts
to study the problem from a more rigorous and different point of
view.

In Refs.~\cite{PKU,piK,crossing}, a new parametrization form --
which we call as the `PKU' parametrization form -- for partial
wave $S$ matrices in the elastic channel is developed,  which,
when combined with chiral symmetry, has been proven  useful in
probing the resonance structure of low energy strong interaction
dynamics. For example, it reveals that the existence of the
$\sigma$~\cite{XZ00,crossing} meson is fully consistent with
chiral symmetry. Combining with crossing symmetry, it further
predicts the $\sigma$ pole mass and width to be $M_\sigma=470\pm
50$MeV, $\Gamma_\sigma=570\pm 50$MeV.~\cite{crossing} Also it is
shown that there should exist the $\kappa$ resonance if the $\pi
K$ scattering length in the I,J=1/2,0 channel does not deviate
much from the value predicted by chiral perturbation
theory.~\cite{piK} The PKU parametrization form is the following,
 \be\label{PKU}
 S^{phy.}=\prod_i S^{poles}_i\cdot S^{cut}\ ,
\ee where $S^{poles}_i$ denote various kinds of poles: resonance,
bound state and virtual bound state. For resonance poles we have
\bqa\label{a pair resonaces S matrix}%
S^R(s) = \frac{{M^2}(z_0)-s + i\rho(s)s G[z_0] }
  {{M^2}(z_0)-s - i\rho(s)s G[z_0]}\ ,
\eqa where
 \bqa {M^2}(z_0) = \mathrm{Re}[z_0] +
\frac{\mathrm{Im}[z_0]\,
     \mathrm{Im}[z_0\,\rho (z_0)]}{\mathrm{Re}[
     z_0\,\rho (z_0)]}\ ,\,\,
G[z_0] =\frac{\mathrm{Im}[z_0]}{\mathrm{Re}[z_0\,\rho (z_0)]}\ ,
\eqa where $z_0$  denotes the resonance pole location on the
complex $s$ plane. The Eq.~(\ref{a pair resonaces S matrix}) is
very interesting as it reveals the remarkable difference between a
narrow resonance located far above the threshold and a light and
broad resonance.
%\begin{theacknowledgments}
%\end{theacknowledgments}
In fact, $s=M^2(z_0)$ is the place where the resonance's
contribution to the phase shift passes $\pi/2$. However, a light
and broad resonance may correspond to a very large $M^2(z_0)$. The
Eq.~(\ref{a pair resonaces S matrix}) for a light and broad pole
actually nicely summarizes the major contribution to IJ=00 channel
$\pi\pi$ scattering phase shift at low energies. The $S^{cut}$ in
Eq.~(\ref{PKU}) no longer contains any pole and for $\pi\pi$
scatterings it can be parameterized as:
 { \bqa\label{f}
 S^{cut}=e^{2i\rho f(s)}\ ,\,\,\,\,\,\,
 f(s)=\frac{s}{\pi}\int_{L}\frac{{\rm
 Im}_Lf(s')}{s'(s'-s)}%\nonumber\\
 +\frac{s}{\pi}\int_{R}\frac{{\rm Im}_Rf(s')}{s'(s'-s)}\
 .
\eqa where $L=(-\infty,0]$ and $R$ denotes physical cuts higher
than the $2\pi$ cut,
%%%%%%%%%%%%%%%%%%%%%%%%%%%%%%%%%%%%%%%%%%%%%%%%%%%%%%%%%%
%%%%%%%%%%%%%%%%%%%%%%%%%%%%%%%%%%%%%%%%%%%%%%%%%%%%%%%%%%
and $\mathrm{Im}_{L,R}f=-\frac{1}{2\rho}\log|S^{phy}|$. Before
proceeding it should be emphasized that the above parametrization
form is only obtainable by assuming analyticity on the whole cut
plane, which can be derived from Mandelstam representation but
nevertheless not proven rigorously from field theory. However the
Lehman--Martin domain of analyticity is
 large enough for phenomenological applications.
Therefore the parametrization form described above may afford a
good approximation to the real situation.

At low energies one may approximate $S^{phy}$ appeared in the
dispersion integral  by $S^{\chi \mathrm{PT}}$ on $L$, to estimate
the background contributions from the left after introducing a
proper cutoff parameter to truncate the dispersion integral. One
may then get more information from the parametrization form
discussed above. Rewrite Eq.~(\ref{PKU}) as
 \be\label{match} \prod_i
S^{poles}_i= S^{phy.}(S^{cut})^{-1} \ ,
 \ee as stated before the $r.h.s.$ of the
above equation can be expressed by low energy quantities appeared
in, for example, the $O(p^4)$ low energy chiral Lagrangian.
Expanding both sides of Eq.~(\ref{match}) at threshold, one
relates the pole parameters to the low energy constants of the
effective Lagrangian. Making use of the $N_c$ counting rule of low
energy constants~\cite{GL85} one can thus trace how the pole moves
on the complex $s$ plane when $N_c$ varies. Nevertheless, in order
to get the pole trajectory we need some further assumptions.
Precisely we assume one pole dominates the $l.h.s.$ of
Eq.~(\ref{match}) for arbitrarily  value of $N_c$. %, and match the
%two sides up to and including $O((s-4m_\pi^2)^2)$ term.
Such an assumption is of course only a speculation and may be
subjected to criticism, though in the case of $N_c=3$  one pole
dominance at low energies is a good approximation.~\cite{crossing}
Nevertheless we will proceed with this working assumption to see
what happens. Here we  keep  all the $N_c$ dependence including
chiral logs. We make use of $O(p^4)$ $\chi$PT results to
approximate $S^{phy.}$ and to calculate $f(s)$, and neglect the
right hand cut integral and truncated the left hand integral at
certain value. In this way we get for $N_c=3$ the pole mass of
$\sigma$ in rather good agreement with more realistic
calculations. The $N_c$ dependence of the pole mass can be traced
numerically. The result is shown in Fig.~\ref{00polefig} and it is
amazing  to see that the pole moves to the negative real axis when
$N_c$ approaches infinity. If the pole location is plotted for
large but finite value of $N_c$ on the $E$ plane, it will stay
somewhere on the complex plane. The latter is the observation made
in Ref.~\cite{palaez} However, the pole trajectory on the $s$
plane clearly indicates what is the correct interpretation: the
mass square becomes negative when the pole moves towards the real
negative axis, rather than that the $\sigma$ resonance has a large
width of $O(1)$. Many phenomenological studies predicted that the
$\sigma$ pole moves towards left, though some finally touch the
negative real axis, some do not.~\cite{ML99}. However it should be
pointed out that the $\sigma$ pole trajectory is very flexible. In
our scheme,  it is actually easy to tune the $L_i$ parameters
within 1$\sigma$ error bar to convert the  pole position to  move
towards the real positive $s$ axis above the threshold.  We have
checked that such a flexibility also exists in the Pad\'e
amplitudes. Therefore our present scheme has a similar prediction
on the pole trajectory comparing with the Pad\'e amplitude.
Therefore only one definite conclusion can be made by us: the
$\sigma$ pole moves to the real axis in the large $N_c$ limit, and
$M^2\sim O(1)$, $G\sim O(1/N_c)$ (more details will appear soon).
Finally we remark that a resonance with negative mass square,
though seems odd, does not seem to spoil any fundamental
principles because it locates on the second sheet. If the
`$\sigma$' discussed in this paper really gets a negative $M^2$,
one wonders wether it has anything to do with the $\sigma$ in the
linear $\sigma$ model.
\begin{figure}%
%\vspace{-1cm}
\includegraphics[height=.2\textheight]{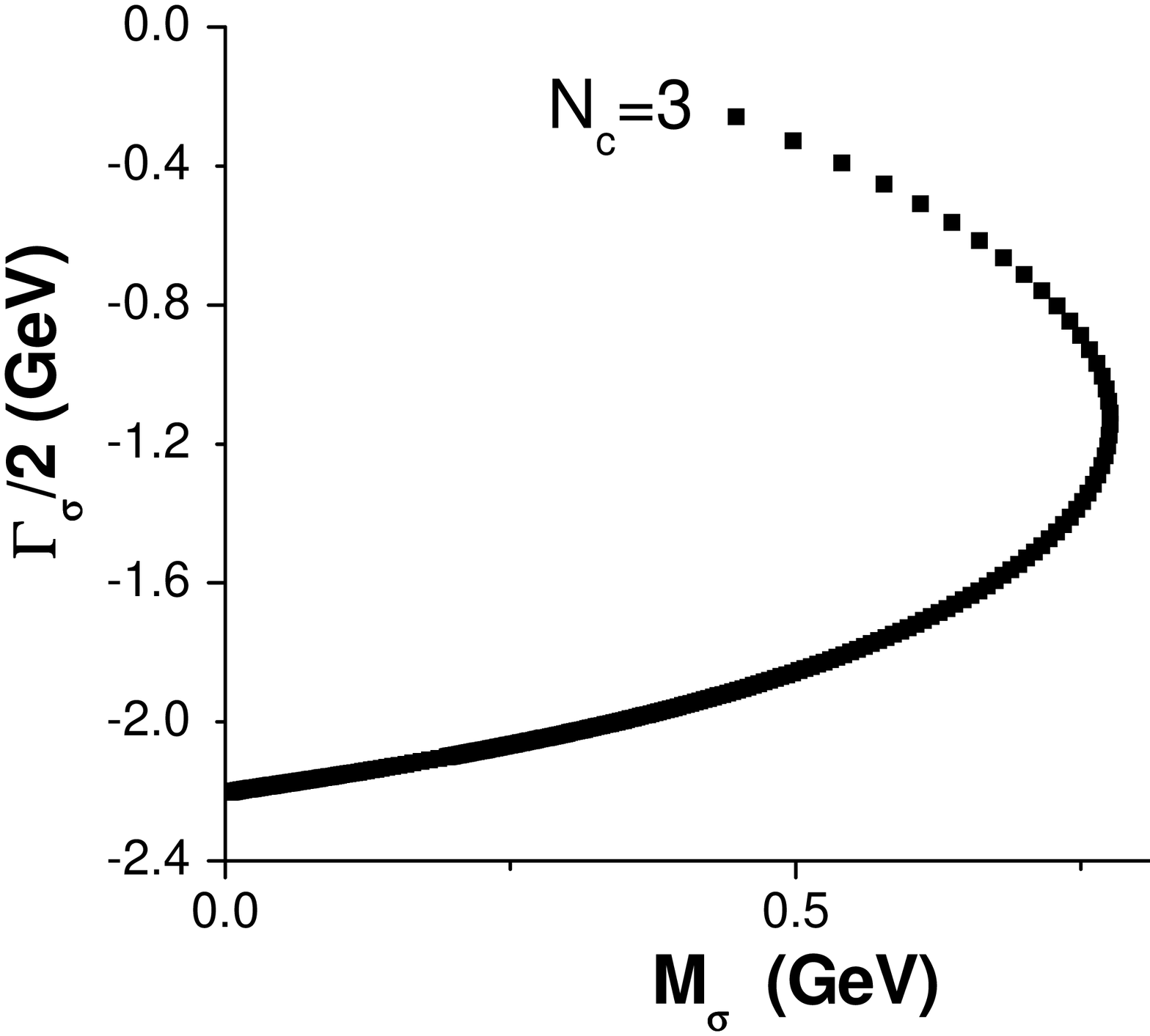}
\includegraphics[height=.2\textheight]{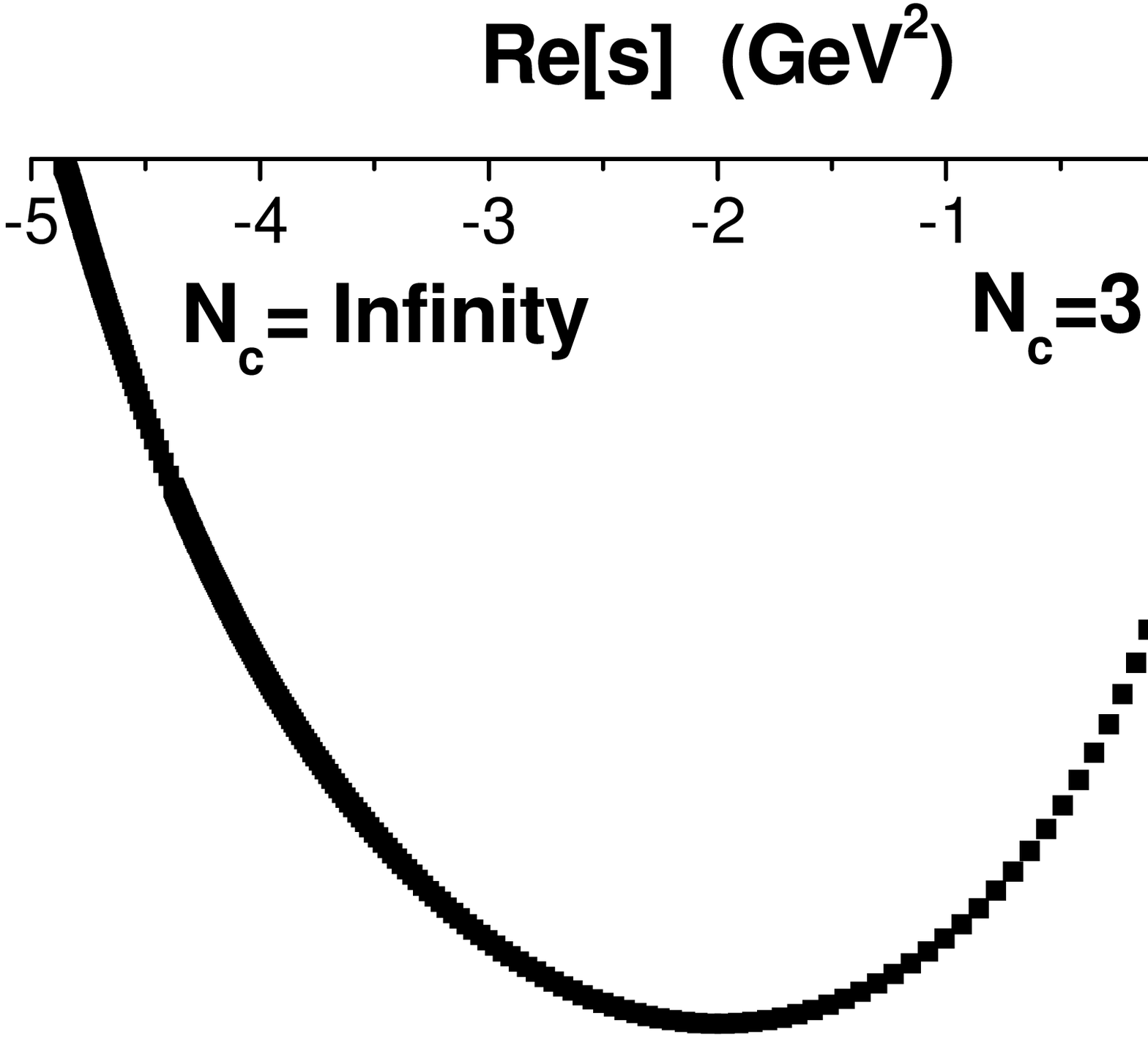}
\vspace{-2cm}
        \caption{The trajectory of the $\sigma$ pole:
        left) the complex $\sqrt{s}$ plane; right) the complex
 $s$ plane.}\label{00polefig}
\end{figure}%

Acknowledgment: We would like to thank Frieder Kleefeld and George
Rupp for  helpful discussions.
%%%%%%%%%%%%%%%%%%%%%%%%%%%%%%%%%%%%%%%%%%%%%%%%%%%%%%%%%%
%%%%%%%%%%%%%%%%%%%%%%%%%%%%%%%%%%%%%%%%%%%%%%%%%%%%%%%%%%%

%\bibitem{kappa}S.~Ishida et al., $Analysis$ $of$ $K\pi$ $scattering$ $phase$
%$shift$ $and$ $the$ $existence$ $of$ $\kappa(900)$
%(hep-ph/9705437)\ ; S.~Cherry and M.~R.~Pennington, $There$ $is$
%$no$ $\kappa(900)$ (hep-ph/0005208)
%\bibitem{ASton} D.~Aston, Nucl. Phys. {\bf B296}(1988)493.
\end{document}